\documentclass[prb,preprintnumbers,twocolumn]{revtex4}
\usepackage{graphicx}
\begin{document}

\markboth{Neutron scattering study of Sn$_2$P$_2$S$_6$}{P. Ondrejkovic et al.}

\title{Neutron scattering study of ferroelectric Sn$_2$P$_2$S$_6$  under pressure}

\author{P. Ondrejkovic$^{\rm a}$}
\author{ M. Kempa$^{\rm a}$}
\author{ Y. Vysochanskii$^{\rm b}$ }
\author{ P. Saint-Gr\'{e}goire$^{\rm c,d}$}
\author{ P. Bourges$^{\rm e}$}
\author{ K. Z. Rushchanskii$^{\rm f}$}
\author{ J. Hlinka$^{\rm a}$}
\affiliation{
$^{\rm a}${\em{Institute of Physics, Academy of Sciences of the Czech Republic, Na Slovance 2, 18221 Praha 8, Czech Republic}}\\
$^{\rm b}${\em{Institute for Solid State Physics and Chemistry, Uzhgorod University 88000, Uzhgorod, Ukraine}}\\
$^{\rm c}${\em{ICGM (UMR CNRS No 5253), C2M, 34095 Montpellier Cedex, France}}\\
$^{\rm d}${\em{Laboratory MIPA, Department of
Sciences and Arts, University of Nimes, 30021 NIMES Cedex 01, France}}\\
$^{\rm e}${\em{Laboratoire L\'{e}on Brillouin, B\^{a}t 563 CEA Saclay, 91191 Gif sur Yvette Cedex, France}}\\
$^{\rm f}${\em{Peter Gr\"{u}nberg Institut, Forschungszentrum
J\"{u}lich GmbH, 52425 J\"{u}lich and JARA-FIT, Germany }}}

\date{\today}

\begin{abstract}
 Ferroelectric phase transition
in the semiconductor Sn$_2$P$_2$S$_6$ single crystal has been
studied by means of neutron scattering in the pressure-temperature
range adjacent to the anticipated tricritical Lifshitz point ($p
\approx 0.18$\,GPa, $T \approx 296$\,K). The observations reveal a
direct ferroelectric-paraelectric phase transition in the whole
investigated pressure range ($0.18-0.6$\,GPa). These results are
 in a clear disagreement with phase diagrams assumed in numerous earlier works, according to
which a hypothetical intermediate incommensurate phase
 extends over several or even tens of degrees in the 0.5\,GPa pressure range.
 Temperature dependence of the anisotropic quasielastic
diffuse scattering suggests that polarization fluctuations present
above $T_{\rm C}$ are strongly reduced in the ordered phase.
Still, the temperature dependence of the ($\bar{2}$00) Bragg
reflection intensity at $p = 0.18$\,GPa
 can be remarkably well modeled assuming the
 order-parameter amplitude growth according to the
power law with logarithmic corrections predicted for a uniaxial
ferroelectric transition at the tricritical Lifshitz point.
\end{abstract}


\maketitle

\section{Introduction} \label{sec_intro}

\begin{figure}
\begin{center}
\includegraphics[width=8cm]{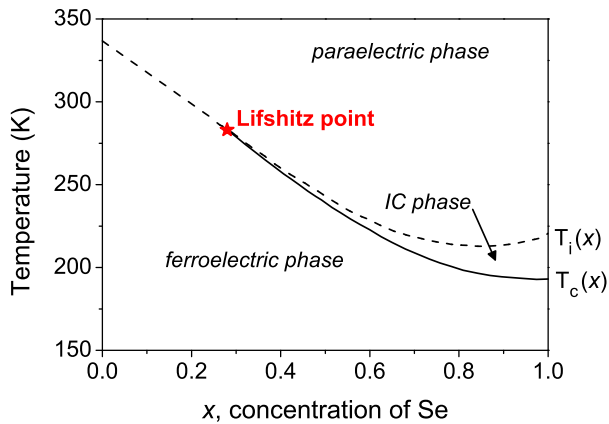}
\caption{(Color online) Schematic temperature-composition phase
diagram for the system Sn$_2$P$_2$(Se$_x$S$_{1-x}$)$_6$ according
to Refs.\,\onlinecite{book, AsiGom}. Dashed line -- second-order
phase transition; solid line -- first-order phase transition. }
\label{newLabelF1Concentration}
\end{center}
\end{figure}

Ferroelectric substances with a narrow electronic band gap have
recently attracted considerable attention due to their potentially
interesting thermoelectric,\cite{lees12}
photovoltaic,\cite{seid10,seid12} and other photoactive
properties.\cite{benn12A, dand12, imla11, tiwa09} Uniaxial
ferroelectric chalcogenides represent one of the best known
families of ferroelectric semiconductors.\cite{benn12B} In
particular, solid solutions of Sn$_2$P$_2$S$_6$ and
Sn$_2$P$_2$Se$_6$ have been recognized as extremely interesting
model systems.
\cite{book, Levanyuk, EijtSnPS, Diehl, moro10, Kosta07, Oleaga,
Kohutych, Korda, AsiGom} All of them have an identical parent
paraelectric structure at high temperatures ($P2_1/n$) and the
same ferroelectric phase at low temperatures
($Pn$).\cite{Carpentier, Barsamian86, Israel} At ambient pressure,
crystals with low Se/S concentration ratio show a direct
ferroelectric-paraelectric phase transition, while those of high
Se/S concentration exhibit an intermediate incommensurate (IC)
phase (see Fig.\,\ref{newLabelF1Concentration}). The IC phase evidenced in Se-rich
solutions\cite{book, Barsamian86} resembles a periodic array of
anti-parallel ferroelectric layers with thickness of about 4\,nm
($|q_{\rm i}|\approx 0.08$\,\AA$^{-1}$). The lower-temperature
phase transition in Sn$_2$P$_2$Se$_6$ could be thus narrated as a
discontinuous transformation from a strictly regular
"ferroelectric nanodomain" arrangement to a structure with usual,
macroscopic ferroelectric domains. This situation is known from
many other so-called type-II IC ferroelectric systems, such as
thiourea, BCCD, NaNO$_2$, etc.\cite{Levanyuk} The peculiarity of
the Sn$_2$P$_2$(Se$_x$S$_{1- x}$)$_6$ solid solution is that here
with decreasing value of Se/S concentration $x$, the temperature
range of the intermediate IC phase continuously decreases and
finally the paraelectric-IC and IC-ferroelectric phase transition
lines join near the so-called Lifshitz point (LP, see
Fig.\,\ref{newLabelF1Concentration}) at $x \approx 0.28$ and $T
\approx 280$\,K.
 Theoretically,  the modulation period should diverge there.\cite{Folk93,Folk99,VysochJETFbis}

Later, it was predicted that a similar LP could also occur in the
temperature-pressure ($T-p$) phase diagram of the pure
Sn$_2$P$_2$S$_6$ crystal. Several experiments indeed confirmed
that the phase transition line splits into two lines after
crossing the point with coordinates around $p=0.18$\,GPa,
$T=296$\,K. The pure Sn$_2$P$_2$S$_6$ system is obviously even
more attractive than the $x=0.28$ compound where, in principle,
the LP-related phenomena could be obscured by compositional
fluctuations inherent to solid solutions.
 In addition, thermodynamic analysis of various physical
quantities indicated that the Landau free-energy expansion
coefficient of the fourth-order term in polarization strongly
decreases with pressure in Sn$_2$P$_2$S$_6$.\cite{book} In fact,
the thermodynamic coordinates of the tricritical point (TCP), at
which this quadratic term vanishes, were predicted to fall in the
vicinity of the above discussed LP. This suggests that the
fortunate intersection of two special lines - the line of the TCPs
and the line of LPs, known also as a tricritical Lifshitz point
(TCLP) - might be reached in Sn$_2$P$_2$S$_6$ under hydrostatic
pressure.\cite{TCLP}

 However, several conflicting statements were reported
about the nature of the high-pressure phase transition of Sn$_2$P$_2$S$_6$.
 In particular, recent analysis  of the specific heat anomalies\cite{Anderson09}
 supports the earlier guess\cite{Sli90} that the conditions for the LP are met at lower pressure than the conditions
for the TCP. This means that only the LP is realized in the pressure-temperature phase diagram of Sn$_2$P$_2$S$_6$.
This scenario was repeatedly assumed in
Refs.\,\onlinecite{Sli99,Guranich01,Guranich05,Gerzanich08,look}
and implies that the special point with coordinates around
$p=0.18$\,GPa, $T=296$\,K is of the same nature as the LP in the
temperature -- concentration diagram of Sn$_2$P$_2$(Se$_x$S$_{1-
x}$)$_6$ solid solution (see Fig.\,\ref{newLabelF1Concentration}). On the contrary, other
recent measurements, such as birefringence measurements of
Ref.\,\onlinecite{Zapeka11} were used to support the original
scenario of the canonical theory of Ref.\,\onlinecite{Vysoch89}:
That the observed special point in the $T-p$ diagram of
Sn$_2$P$_2$S$_6$ is merely a TCP but surely not a
LP.\cite{Zapeka11} Moreover, the absence of the LP in the $T-p$
phase diagram of Sn$_2$P$_2$S$_6$ was also inferred from the
comparisons with the Blume-Emery-Griffith's model.\cite{Vysoch10}
On the top of that, there is a considerable disagreement about the
domain of stability of the high-pressure intermediate phase of
Sn$_2$P$_2$S$_6$ among various
sources.\cite{Sli99,Tyagur,Anderson09,Say}

Despite these challenging issues, diffraction investigations of
the intermediate phase of Sn$_2$P$_2$S$_6$ were not reported so
far and there is no direct evidence of its IC nature. The aim of
this work is to bring at least a partial answer to the above
raised issues by discussion of new results from high-pressure
investigations of Sn$_2$P$_2$S$_6$ by means of neutron scattering.

\section{Experimental details} \label{sec_exp}

The neutron scattering experiment was performed at the Laboratoire
L\'{e}on Brillouin (Saclay, France) on the 2T1 thermal neutron beam
and 4F2 cold neutron beam three-axis spectrometers (TASs).
Spectrometers were operated with PG(002) analyzer set to the
neutron wave number $k_{f}=$ 2.662\,\AA$^{-1}$ and $k_{f}=$
1.48\,\AA$^{-1}$, respectively. The energy resolution at
zero-energy transfer was 0.2 and 0.05 THz, the Q resolution about
0.01\,\AA$^{-1}$ and 0.005\,\AA$^{-1}$ on the 2T1 and 4F2 TASs,
respectively.

 The  Sn$_2$P$_2$S$_6$ single crystal used in this experiment was
 grown from melt by the Bridgeman method at Uzhgorod University.
The sample ($\approx0.8$\,cm$^{3}$) was placed in a special helium
pressure cell designed for neutron scattering experiments in a
temperature range of
 $20-305$\,K and under hydrostatic pressures up to about 0.6\,GPa.

\begin{figure}
\begin{center}
\includegraphics[width=8cm]{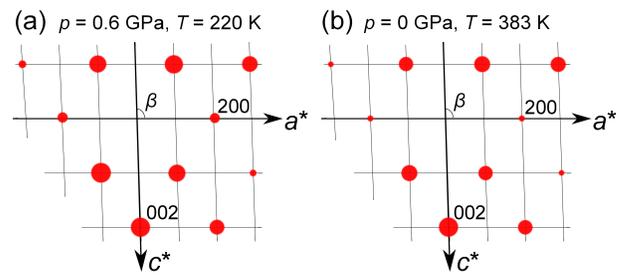}
\caption{(Color online) Accessible part of the $a$*$c$* scattering
plane with schematic representation of neutron scattering
intensities of principal Bragg reflections in the paraelectric
phase. The areas of the circles are proportional to absolute
values of neutron scattering structure factors
 (a) in the present experiment at $p=0.6$\,GPa and $T=220$\,K and (b) $p=0$\,GPa and $T=383$\,K values recalculated from the x-ray data of Ref.\,\onlinecite{para} using the Jana2006 computer program\cite{JANA} with the
neutron wave number set to $k_{f}=$ 1.48\,\AA$^{-1}$. }
\label{newLabelF2Map}
\end{center}
\end{figure}

The crystal was mounted so that the scattering plane matched its
crystallographic mirror plane $m_{y}$ (throughout the paper, we
stick to the pseudo-orthorhombic unit cell defined by $a=9.378$\,\AA,
$b=7.488$\,\AA, $c=6.513$\,\AA, and $\beta=91.15^\circ$, as in
Ref.\,\onlinecite{ferro}).
 This
horizontal plane thus also contains ferroelectric polarization
direction (at about $10-15^\circ$ out of the $a$ axis). A
reasonably accessible part of the reciprocal space in such a
scattering geometry is shown in Fig.\,\ref{newLabelF2Map}.
 A rough survey of uncorrected
neutron diffraction intensities of main Bragg reflections in
Sn$_2$P$_2$S$_6$, as detected at $T=220$\,K and $p=0.6$\,GPa in
our experiment, is indicated by variably sized symbols in
Fig.\,\ref{newLabelF2Map}(a). Results agree fairly well with the
neutron diffraction intensities calculated from structural data of
the ambient-pressure paraelectric phase of Sn$_2$P$_2$S$_6$
[Fig.\,\ref{newLabelF2Map}(b), data at $T=383$\,K of
Ref.\,\onlinecite{para}].

The $a^*c^*$ plane of Fig.\,\ref{newLabelF2Map} is the same $a^*c^*$ plane within
which the satellite reflections with a modulation wave vector ${\bf q}_{\rm i}\approx 0.085 {\bf c^*} - 0.01 {\bf a^*}$,
 demonstrating the IC
phase, were observed in the
Sn$_2$P$_2$Se$_6$ crystal.\cite{Barsamian86}
 Independent x-ray and neutron scattering experiments\cite{EijtSnPSe,Barsamian93} agree
that satellite reflections of Sn$_2$P$_2$Se$_6$ are particularly
strong for large momentum transfers $\bf Q$, roughly parallel to
the spontaneous polarization (${\bf Q} \parallel {\bf P}_{\rm
s}$). This is a consequence of the usual situation caused by the
transverse character of the frozen polarization wave (${\bf
q}_{\rm i} \bot {\bf P}_{\rm s}$), and the same can be anticipated
for the supposed Sn$_2$P$_2$S$_6$
  satellite reflections.
  Clearly,
detection of satellites with a very small modulation wave vector,
comparable to the tails of the $Q$ resolution of the instrument,
could be problematic in the vicinity of a strong Bragg reflection.
  Unfortunately, systematic absences  are
common to both main and satellite reflections, that is, the
satellites are expected only around the allowed reflections in the
$a^*c^*$ plane. For this reason, most of our measurements were
taken in the vicinity of the $(\bar{2}00)$ Bragg reflection. This
is an allowed reflection, but rather a weak one, and at the same
time the corresponding scattering vector ${\bf Q}$ is almost
parallel to ${\bf P}_{\rm s}$.

 \section{Results}

\begin{figure}
\begin{center}
\includegraphics[width=8cm]{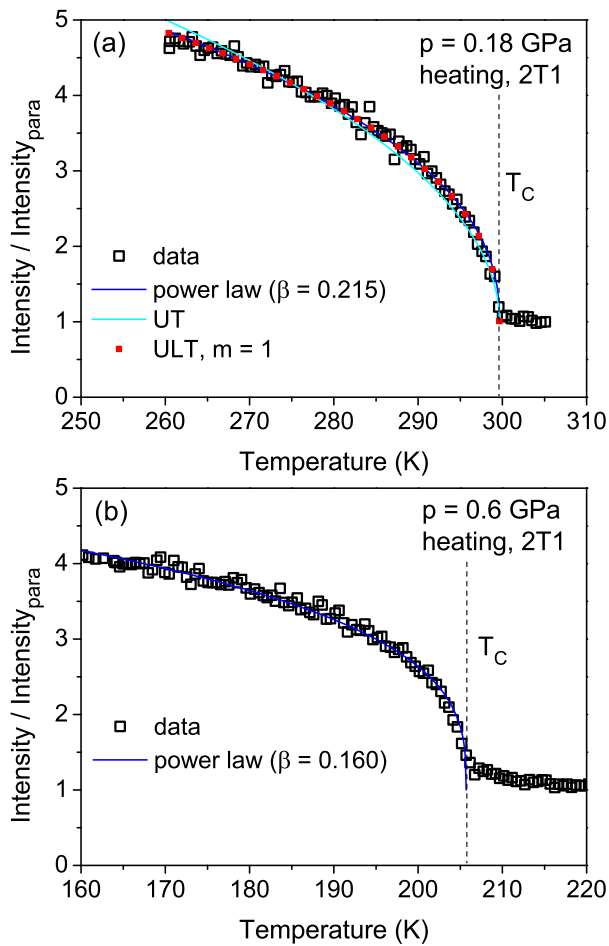}
\caption{(Color online) Temperature dependence of the
$(\bar{2}00)$ reflection intensity (a) at 0.18\,GPa and (b) at
0.6\,GPa. The full points and solid lines refer to the fits using
Eqs. (\ref{intensity}) and (\ref{order parameter}) with and
without logarithmic corrections, respectively.} \label{newLabelF3}
\end{center}
\end{figure}

In agreement with our anticipation, the neutron diffraction
intensity of the $(\bar{2}00)$ Bragg reflection in the
 paraelectric phase was several times smaller than in
the ferroelectric phase at ambient conditions. In fact, monitoring
of the $(\bar{2}00)$ reflection allowed us to probe quite
accurately the ferroelectric order parameter. As an example, the
temperature dependence of the $(\bar{2}00)$ Bragg reflection
intensity at 0.18\,GPa and 0.6\,GPa is shown in
Figs.\,\ref{newLabelF3}(a) and \ref{newLabelF3}(b). The intensity
visibly starts to rise at a well defined temperature that can be
associated with the ferroelectric phase transition. The
temperatures thus obtained are given in Table\,I and within the
possible experimental error in the pressure and temperature
determination they correspond well to those derived from the
compressibility anomalies reported in Ref.\,\onlinecite{Sli99}.

In the vicinity of these transition temperatures, we have
 searched for possible indications of a satellite
reflection. The pressure of 0.18\,GPa is supposedly that of the
TCLP, while the measurements at 0.3\,GPa and 0.6\,GPa are taken in
the region where finite temperature range of the IC phase was
anticipated. For example, we have recorded intensity maps in a 2D
area around the $(\bar{2}00)$ Bragg reflection, extending up to
$\pm 0.2$\,r.l.u. along both $a^*$ and $c^*$ at 0.6\,GPa and
temperatures of 205.5, 206, and 206.5\,K (at 0.2, 0.7, and 1.2\,K
above $T_{\rm C}=205.3$\,K) and similar maps were recorded also at
0.3\,GPa and temperatures 267, 270, 273, and\,276 K ($T_{\rm
C}=271$\,K). However, no obvious satellite peaks have been seen.

\begin{figure}
\begin{center}
\includegraphics[width=8cm]{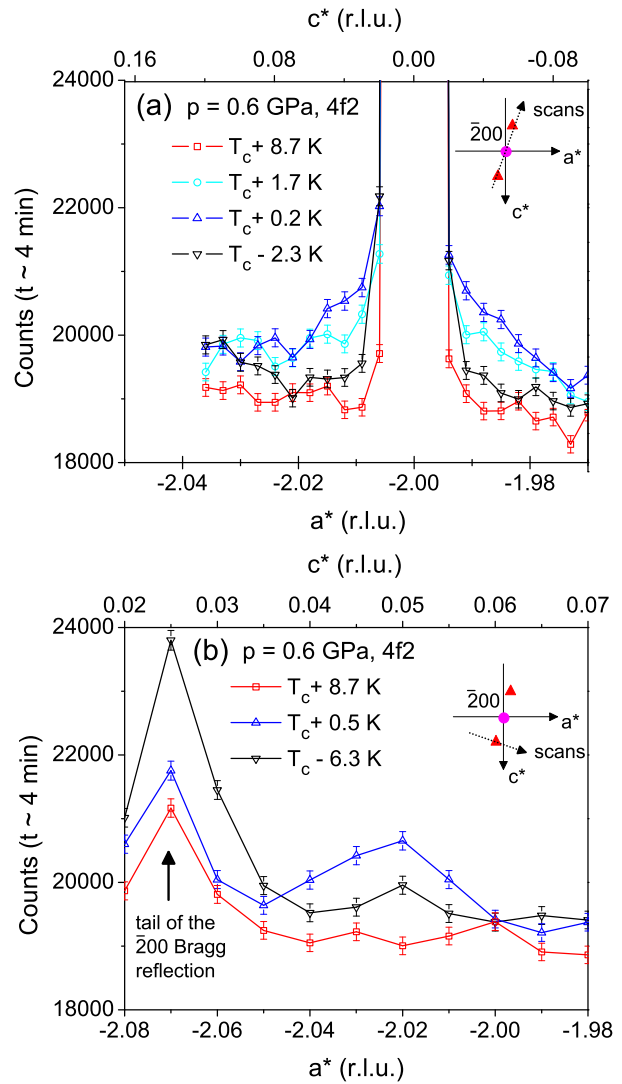}
\caption{(Color online) Zero-energy-transfer scans in the
$(\bar{2}00)$ Brillouin zone of the Sn$_2$P$_2$S$_6$ crystal taken
 in the vicinity of the ferroelectric phase
 transition (at 0.6\,GPa; near $T_{\rm C} = 205.3$\,K). (a) Scans
 in the
$ [-0.03,0,1]$ direction (approximately along the diffuse streak),
and (b) scans in the $[1,0,0.05]$ direction (across the diffuse
streak). Measured scan trajectories
 are schematically
 indicated in the insets. Solid lines are just guides
for the eye.} \label{newLabelF4}
\end{center}
\end{figure}

\begin{figure}
\begin{center}
\includegraphics[width=7cm]{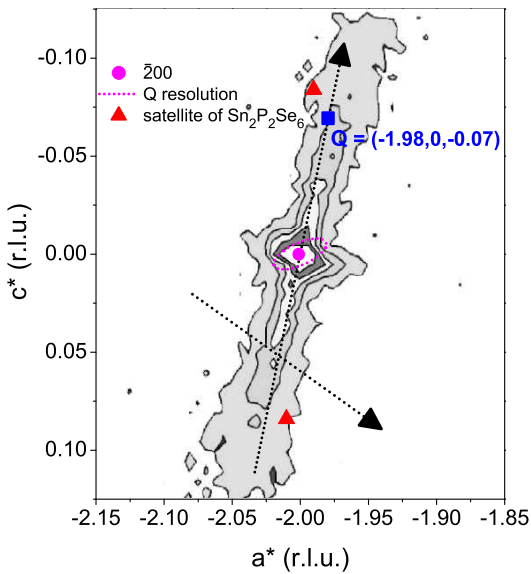}
\caption{(Color online) Schema of reciprocal-space trajectories of
data scans shown in Fig.\,\ref{newLabelF4}, superposed with
contour plots of the critical diffuse scattering known from
ambient pressure x-ray diffraction measurements of
Ref.\,\onlinecite{Hlinka05}. Full triangles indicate positions of
incommensurate satellite reflections of the ambient-pressure
incommensurate phase of the related compound Sn$_2$P$_2$Se$_6$, as
reported in Ref.\,\onlinecite{Barsamian86}. } \label{newLabelF5}
\end{center}
\end{figure}

At the same time, our observations disclosed traces of weak and
broad diffuse scattering ridges, extending over the "soft"
direction, roughly perpendicular to the direction along which the
spontaneous polarization is formed. Figure\,\ref{newLabelF4}(a)
shows an example of a scan taken along this diffuse scattering
ridge at several temperatures. Examples of transverse scans are
shown in Fig.\,\ref{newLabelF4}(b). Trajectories of these scans
are schematically shown in Fig.\,\ref{newLabelF5}, along with the
contour plots of the ambient-pressure diffuse scattering
topography taken from Ref.\,\onlinecite{Hlinka05}. Note that the
diffuse scattering ridge corresponds to the smaller peak in
Fig.\,\ref{newLabelF4}(b), while the larger intensity peak is
merely a spurious leakage of $(\bar{2}00)$ Bragg reflection
scattering due to the strongly anisotropic tails of the
instrumental resolution function.

\begin{figure}
\begin{center}
\includegraphics[width=6cm]{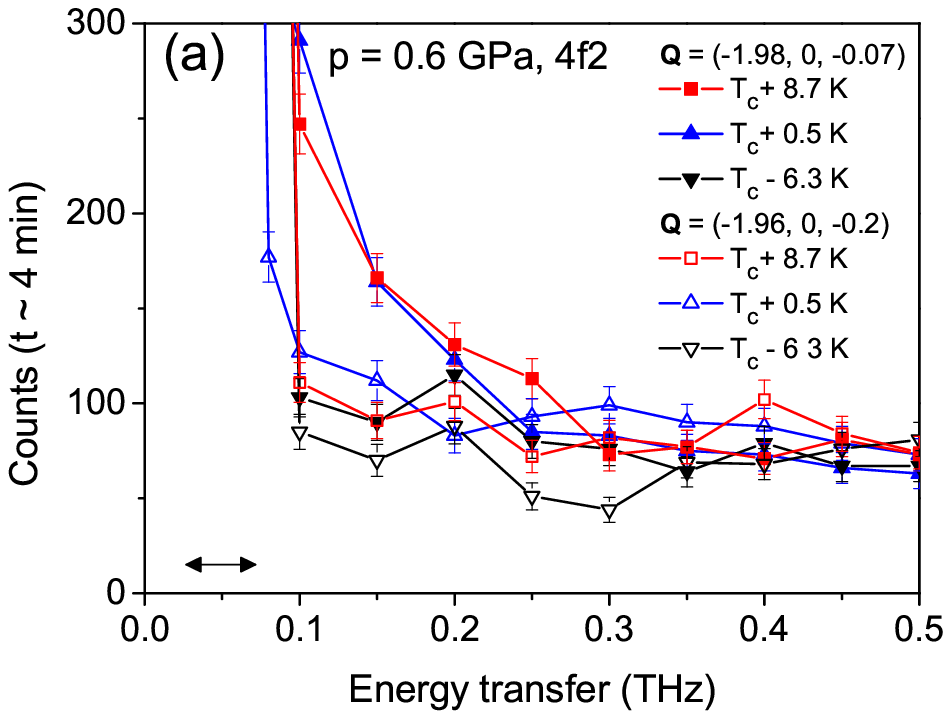}
\includegraphics[width=8cm]{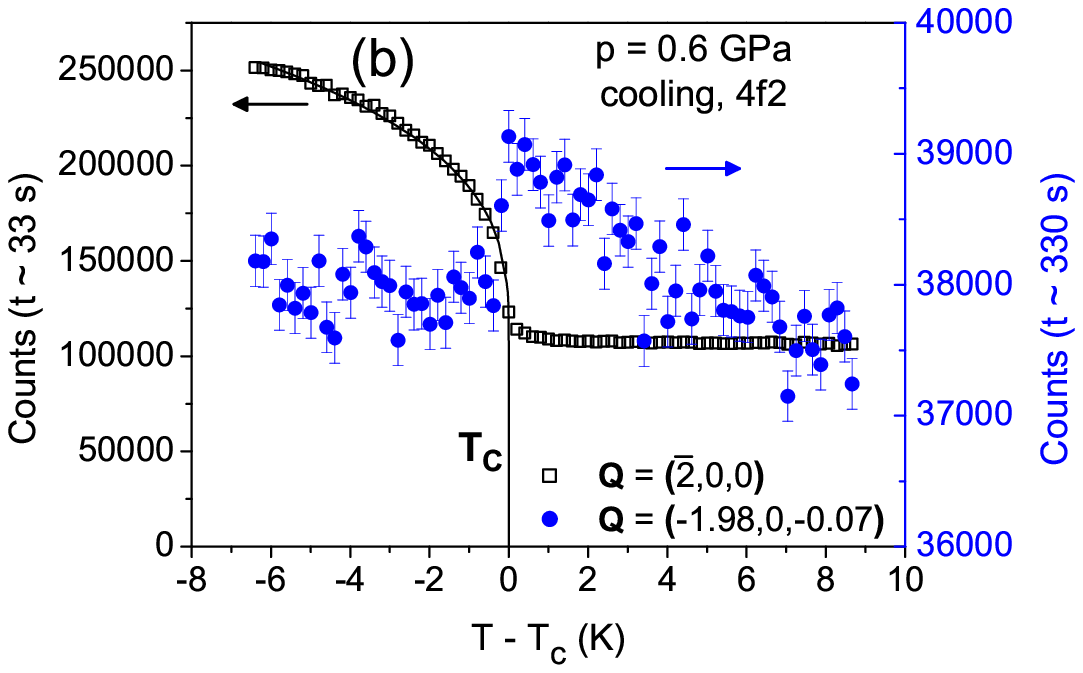}
\caption{(Color online) Temperature dependence of both elastic and quasielastic scattering in the $(\bar{2}00)$ Brillouin zone at 0.6 GPa:\\
(a) selected constant-Q scans  for several temperatures,
(b) principal Bragg reflection intensity at {\bf q} $=(0,0,0)$ and elastic diffuse scattering at {\bf q} $=(0.02,0,-0.07)$.} \label{newLabelF6}
\end{center}
\end{figure}

Some of the scans displayed in Fig.\,\ref{newLabelF4}(a) show a
non-monotonic intensity decay with the distance from the Brillouin
zone center. However, due to the considerable background intensity
caused by high-pressure-cell environment, this can hardly be
considered as a significant indication of incommensurate satellite
reflections. Rather, the anisotropy of this diffuse scattering is
obvious and strikingly similar to the critical fluctuations
previously seen by x-ray scattering at ambient pressure
conditions.\cite{Hlinka05} The assignment of the diffuse
scattering to the order-parameter fluctuations is further
corroborated by the observed increase in its intensity as the
transition is approached from above and by its rather abrupt
reduction just below the phase transition point [see full points
in Fig.\,\ref{newLabelF6}(b)]. Moreover, few constant-${\bf Q}$
spectra taken at selected positions in the high-temperature phase
[Fig.\,\ref{newLabelF6}(a)] suggest that this diffuse scattering
is quasielastic in its nature, with FWHM of the order of 0.2\,THz.
It will be shown in the next section that several important
conclusions can be drawn from such measurements, even though the
high elastic and inelastic scattering background did not allow
their more quantitative analysis.

\section{Discussion}

\begin{figure}
\begin{center}
\includegraphics[width=8.5cm]{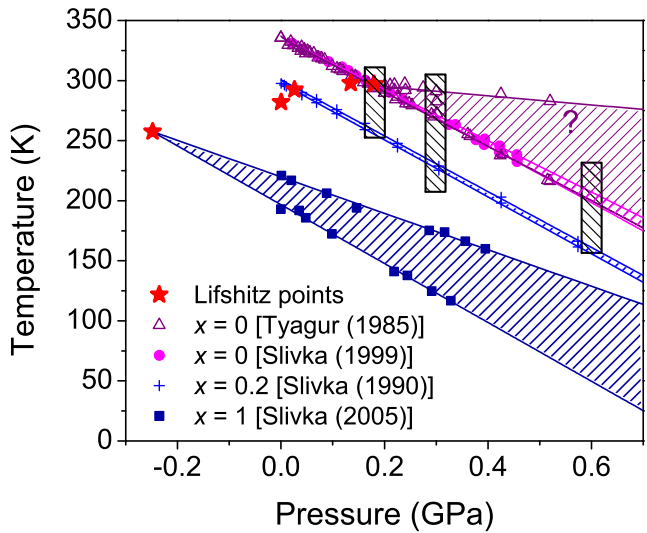}
\caption{(Color online) Temperature-pressure phase diagram of
Sn$_2$P$_2$(Se$_x$S$_{1- x}$)$_6$ solid solution.\cite{Tyagur,
Sli99, Sli90, Slivka05, Barsamian86} The shaded areas indicate
intermediate phases between high-temperature paraelectric and
low-temperature ferroelectric phases for given concentrations $x$.
The stars stand for Lifshitz points of solid solution with $x$
equals: 1, 0.28, 0.2 0.04, and 0 (from left to right). Temperature
ranges of neutron scattering measurements reported in this work
are indicated by vertical rectangular areas.} \label{newLabelF7}
\end{center}
\end{figure}

{\it Absence of the TCLP, three alternative scenarios.}
Summarizing, our results do not provide any support for the LP in
the $T-p$ phase diagram of Sn$_2$P$_2$S$_6$ up to the hydrostatic
pressure values of about 0.6\,GPa. Thus, the most natural
conclusion seems to be that there is no IC phase in this $T-p$
region at all ("scenario A"). The absence of any clear evidence
for the LP in the above displayed observations led us nevertheless
to perform detailed and careful comparisons of our results with
those of previously reported experiments, which did confirm the
presence of the LP. In particular, we have traced the $T-p$ phase
diagram of Sn$_2$P$_2$S$_6$ (see Fig.\,\ref{newLabelF7}) according
to the reported anomalies in the compressibility data
(Ref.\,\onlinecite{Sli99}, circles in Fig.\,\ref{newLabelF7}) and
in the dielectric permittivity data (Ref.\,\onlinecite{Tyagur},
triangles in Fig.\,\ref{newLabelF7}). Both these measurements
suggest splitting of the phase transition in Sn$_2$P$_2$S$_6$
($x=0$) at around $p=0.18$\,GPa, $T=296$\,K, which was naturally
interpreted as the expected LP. At the same time, the stability
range of the anticipated intermediate phase is markedly different
in these two measurements: It opens at the rate of about
$190$\,K/GPa according to Ref.\,\onlinecite{Tyagur} ("scenario
B"), while there is only about
 $20$\,K/GPa opening rate deduced from measurements of Ref.\,\onlinecite{Sli99} ("scenario C").

It is also instructive to compare these data with the
$T-p$ phase diagram of the Sn$_2$P$_2$Se$_6$ crystal, in which the
 IC phase was firmly established at
least at ambient pressure. In fact, the temperature range of the
IC phase of Sn$_2$P$_2$Se$_6$ (limited by pairs of square symbols
in Fig.\,\ref{newLabelF7}) opens at a rate of about $100$\,K/GPa,
which is just an intermediate value between the two visibly
conflicting values given for Sn$_2$P$_2$S$_6$ (20 and 190 K/GPa,
respectively). Let us note that the similar data for the $x=0.2$
solid solution crystals (cross symbols, Ref.\,\onlinecite{Sli90})
and the $x=0.05$ solid solution crystals
(Ref.\,\onlinecite{Sli90}, not included in
Fig.\,\ref{newLabelF7}), suggest already a quite narrow range of
the IC phase, which seems to be much more consistent with the
narrow IC phase for Sn$_2$P$_2$S$_6$ (scenario C). The latter
tendency could be also easily associated with the fact that the IC
phase temperature range width is strongly decreasing with the S/Se
ratio at ambient pressure (see Fig.\,\ref{newLabelF1Concentration}). Nevertheless, the phase
stability of the IC phase is a strongly nonlinear function of the
Se concentration at ambient pressure (see
Fig.\,\ref{newLabelF1Concentration}) and so it could be a
non-monotonous function of the Se concentration at higher
pressures.

In spite of these disagreements about the temperature range of the
IC phase in Sn$_2$P$_2$S$_6$, results of Ref.\,\onlinecite{Sli99}
(scenario C) and
Refs.\,\onlinecite{Tyagur},\,\onlinecite{Anderson11} (scenario
B) agree well about its onset at about $0.18$\,GPa, $T=296$\,K.
This location of the LP appears to be fairly well consistent with
$T-p$ coordinates of LPs reported for the several solid solutions
with different Se concentrations (LPs are denoted by star symbols
in Fig.\,\ref{newLabelF7}). Likewise, the "theoretical" LP of the
Sn$_2$P$_2$Se$_6$ compound, obtained by extrapolating of the
experimental paraelectric-IC and the IC-ferroelectic phase
transition temperatures towards negative pressure values (see
Fig.\,\ref{newLabelF7}), seems to lie on a more or less straight
line in the projected $T-p$ diagram.

{\it Elimination of the scenario B by continuity arguments.} So
how credible is the outcome of the present result with respect to
the previous experiments? Perhaps the most significant advantage
of the present experiment is that here we employed a scattering
technique, which allows us to appreciate the amplitude of the
order parameter as well as its wave vector. In particular, it is
important that our experiments positively prove that the
zone-center ferroelectric order parameter vanishes at temperatures
given in Table\,I in a fairly continuous manner. It is well known
that virtually all displacement-type IC dielectrics obey the
standard continuity paradigm supported by the Landau
theory\cite{Levanyuk}: (i) the phase transition from the
paraelectric to the IC phase is a continuous transition, and (ii)
 the modulated-ferroelectric "lock-in" transition is
  a first-order phase transition but with a relatively moderate jump of the modulation amplitude.
Thus, already from these results, there is no space for an IC
phase above temperatures given in Table\,I. Based on this
argument, we can safely refute the phase diagram with a broadly
opened IC phase\cite{Anderson11,Tyagur} (scenario B).

In order to make our conclusion even more convincing, let us
recall that the atomic positions ${\bf r}_j$ in the vicinity of
the ferroelectric structural transition can be approximately
expressed through the paraelectric reference ones (${\bf r}_{j,\rm
para}$) by means of the frozen order-parameter amplitude $\eta$
and corresponding eigenvector ${\bf e}_j$ as
\begin{eqnarray}
{\bf r}_j ={\bf r}_{j,\rm para}  + \eta {\bf e}_j ~.
\end{eqnarray}
Inserting this into the standard expression for the elastic
neutron scattering structure factor
\begin{eqnarray}
F({\bf Q})= \sum\limits_{j} b_j e^{i {\bf Q}\cdot{\bf r}_j}
e^{-W_j} ~,
\end{eqnarray}
where $b_j$ is the neutron scattering length, $W_j$ is the
Debye–-Waller factor of the $j$th atom, and $\bf{Q}$ is the
scattering vector and keeping the leading term in $\eta$, one
obtains the reliable and frequently exploited relation
\begin{eqnarray}
F(\eta, {\bf Q}) \approx F(0, {\bf Q})( 1 + i \eta {\bf Q}\cdot {\bf
e}_j).
\end{eqnarray}
This expression implies that the order parameter is modifying the
elastic coherent neutron scattering intensity as
\begin{eqnarray}
I(\eta,{\bf Q})&=& F(\eta, {\bf Q}). F(\eta, {\bf Q})^* \nonumber \\
&=& I(0,{\bf Q})+\eta^2 I(0,{\bf Q}) \left| {\bf Q}\cdot{\bf e}_j
\right|^2 ~. \label{intensity}
\end{eqnarray}
Thus, the extraneous contribution to the Bragg reflection
intensity is proportional to the square of the ferroelectric order-parameter amplitude ($\eta^2$). Moreover, it can be shown, that
for a long-wavelength IC phase of type II, the second term in the
above expression also quite well approximates the satellite
reflection intensity. Consequently,
 the intensity by which the principal reflection is
increased just below the lock-in transition should be comparable
to the intensity of the satellite reflection just above the
lock-in transition.

In other words, one may expect that above the lock-in phase
transition, the satellite reflection in the vicinity of the
$(\bar{2}00)$ Bragg reflection should have about the same
intensity as the order-parameter-related part of the $(\bar{2}00)$
Bragg reflection intensity itself. Here in our measurements (see
Fig.\,\ref{newLabelF3}) the anomalous intensity of the
$(\bar{2}00)$ Bragg reflection decreases considerably and {\em
continuously} as the temperature approaches the transition
temperature (given in Table\,I). from below, and thus we know that
the order parameter vanishes there completely. This suggests that
such transition cannot be a lock-in transition, and a modulated
phase persisting above this temperature as proposed from the
diagram of Refs.\,\onlinecite{Anderson11,Tyagur} (scenario B)
can hardly be expected.

{\it Comparison of scenarios of A and C.}
 Therefore, the only possibility on how to reconcile our
observation with a hypothesis of an IC phase is to assume that it
exists {\em below} the transition temperatures of Table\,I. This
could happen for a modulation wave vector smaller than the
momentum resolution, as then the satellite reflection intensity
would be integrated into the $(\bar{2}00)$ Bragg reflection
scattering. Since a systematic broadening of the $(\bar{2}00)$
Bragg reflection was not noticed in our experiments, it would mean
that the modulation period is of the order of 100\,nm. The phase
transition temperature of Table\,I would correspond to a
paraelectric-IC phase transition $T_{\rm i}$. This is in an
acceptable agreement with phase transition temperatures given by
Ref.\,\onlinecite{Sli99} (scenario C).

 From the same work,\cite{Sli99} one may infer that the lock-in transitions $T_{\rm lock-in}$ would
fall at about 3\,K (5\,K) below $T_{\rm i}$ at 0.3\,GPa (at
0.6\,GPa). Obviously, a small jump of the order-parameter
amplitude at these temperatures can hardly be identified in our
data. In principle, this scenario seems to be compatible also with
the birefringence data in Ref.\,\onlinecite{Say}. There are indeed
tiny jumps in the temperature dependence of the anomalous part of
the birefringence of Sn$_2$P$_2$S$_6$ at high pressures, even
though the data alone were not considered as
significant.\cite{Say} At the same time, very similar jumps were
observed at the well established $T_{\rm lock-in}$ temperature in
the temperature dependence of the birefringence data of
Sn$_2$P$_2$Se$_6$, and they are also quite
small,\cite{UFZ1,UFZ2,UFZ3} despite a much broader range of the IC
phase there and presumably more abrupt lock-in transition. Let us
note that in contrast, the compressibility data of
Ref.\,\onlinecite{Sli99} show a larger anomaly at the presumed
$T_{\rm lock-in}$ transition, while $T_{\rm i}$ is barely
noticeable. The compressibility anomalies, however, do not
directly scale with the order-parameter amplitude. On the
contrary, it is likely that the compressibility anomalies are
sensitive to the modulation period, similarly as it was
demonstrated for example in the case of modulated phases of
BCCD.\cite{HlinkaBCCD} Therefore, the large compressibility
anomaly at $T_{\rm lock-in}$ might be compatible with scenario C
as well.

    Several possible experimental problems that might be at the origin
of the contradictory reports on the splitting of the phase
transition in Sn$_2$P$_2$S$_6$ have been recently discussed for
example in Ref.\,\onlinecite{Say}. Perhaps the most serious is the
peculiar sensitivity of Sn$_2$P$_2$(Se$_x$S$_{1-x}$)$_6$ solid
solutions to the absorption of light. It was reported that an
$x=0.2$ crystal irradiated by 0.5\,W krypton (647\,nm) laser
radiation shows two close anomalies in the temperature dependence
of the light transmission even though the crystal with this
composition has nominally a single continuous phase
transition.\cite{VysochJETF} We speculate that this partly
understood photovoltaic or photoelastic phenomenon could be
responsible for the extraneous anomalies that are behind "scenario
B" (experiments reported in Ref.\,\onlinecite{Tyagur}). However,
strictly speaking, same worries apply to the experiments
supporting the narrow IC phase (scenario C).

   \begin{table}
\caption{Phase transition temperatures and $\beta$ parameters of formulas
(\ref{intensity}, \ref{order parameter})  without any
logarithmic correction for the $(\bar{2}00)$ Bragg reflection
measured at given pressure $p$ in the ferroelectric phase on the
2T1 TAS. }
    {\begin{tabular}{@{}lccc}
   $p$ (GPa)      & $T_{\rm C}$ (K) & $\beta$   & heating rate (K/min)   \\
    \colrule

   0.18         & $299.6$   & $0.215$     & $0.22$     \\
   0.3          & $271.0$   & $0.196$     & $0.06$      \\
   0.6          & $205.3$   & $0.160$     & $0.22$      \\
    \end{tabular}}
\label{parameters}
\end{table}
\begin{table}
\caption{List of critical dimensions d$_{\rm u}$ and critical
exponents for uniaxial ferroelectrics according to Refs.
\onlinecite{Folk93,Folk99}. The star symbol
* indicates the logarithmic corrections in the power laws.
(U-uniaxial dipolar, L - Lifshitz, T - tricritical, $m$-dimensions
of Lifshitz subspace).}
    {\begin{tabular}{@{}llllccc}
   system & d$_{\rm u}$ & $\alpha$       & $\beta$ & $\gamma$  & $x$ &   \\
   \hline
   U              & 3   & 0*             & $\frac{1}{2}$*    & 1*   & $\frac{1}{3}$     \\
   T              & 3   & $\frac{1}{2}$* & $\frac{1}{4}$*    & 1*   & $\frac{1}{3}$    \\
   UT             & $2\frac{1}{2}$  & $\frac{1}{2}$  & $\frac{1}{4}$     & 1    & -      \\
   ULT, $m=1$   & 3     & $\frac{1}{2}$* & $\frac{1}{4}$*      & 1*   & $\frac{1}{10}$    \\
    \end{tabular}}
\label{models}
\end{table}

{\it Order-Parameter fluctuations.} Additional support for
scenario A can be derived from the observations of the
temperature dependence of the critical fluctuations of the order
parameter. In displacive ferroelectrics, the intensity of the
associated quasielastic diffuse scattering is typically
proportional to $1/\omega_{\rm SM}^4$, where $\omega_{\rm SM}$ is
the soft mode frequency at a given wavevector. Therefore, the
increase of the critical scattering in the vicinity of the phase
transition is directly associated with soft branch frequency
lowering. In the incommensurate phase, the soft mode branch is
split into two branches, called phason branch and amplitudon
branch, respectively.\cite{Levanyuk,Currat} Obviously, the diffuse
scattering is mainly determined by the lower frequency phason
branch.\cite{thiourea} Since for small modulation amplitudes the
phason gap remains zero, the phason branch and the diffuse
scattering does not show much temperature evolution within the
range of incommensurate phase.\cite{thiourea,Levanyuk,Currat}

The critical quasielastic diffuse scattering observed within a few
degrees above the phase transition in present experiments (at
0.6\,GPa, see Figs.\,\ref{newLabelF4} and\,\ref{newLabelF6}) can
be indeed well understood as due to an overdamped soft mode
branch. In particular, the pronounced temperature dependence of
the diffuse scattering intensity shown in
Fig.\,\ref{newLabelF6}(b) (data corresponding to the right-hand
side scale) suggests soft mode frequency lowering upon approaching
the nominal transition temperature from above. Such temperature
dependence is typical for a paraelectric phase, but not for an
incommensurate phase. At the same time, the steep intensity drop
below $T_{\rm C}$ suggests marked stiffening of the soft mode
frequency. This can be indeed expected for a ferroelectric phase
transition with a partial order-disorder character (see discussion
in Refs.\,\onlinecite{Dvorak,Padlewski}) and it is in agreement
with direct measurements of both the static dielectric
permittivity temperature dependence
 and the soft-mode frequency temperature
dependence (Ref.\,\onlinecite{Volkov}) near ambient-pressure
phase transition of Sn$_2$P$_2$S$_6$.

As already mentioned, the quasielastic, that is, dynamical nature
of this diffuse scattering intensity is apparent from
Fig.\,\ref{newLabelF6}(a). Spectra taken in the immediate vicinity
of $T_{\rm C}$ suggest a bell-shaped quasi-elastic intensity
component with FWHM of the order of 0.2\,THz, that is, visibly
exceeding the instrumental energy resolution. Dynamic critical
fluctuations with relaxational frequencies of the order of
0.2\,THz would fit quite well with typical relaxation frequencies
expected for soft mode fluctuations in the system (for an
overdamped soft phonon mode, the relaxational frequency is about
$\omega_{\rm SM}^2/\gamma_{\rm SM}$, where $\gamma_{\rm SM}$ is
the damping of the soft phonon mode.) Moreover, the quasielastic
signal which for temperatures near $T_{\rm C}$ would reach the
values of the order of 1000 counts at zero energy transfer in
Fig.\,\ref{newLabelF6}(a) (extrapolation of full symbols to
zero-energy) does match quite nicely the 200 counts per minute of
the q-dependent and T-dependent diffuse signal apparent from
Figs.\,\ref{newLabelF4} and\,\ref{newLabelF6}(b). Thus, in brief,
all the observations considered here are in a rather good
agreement with the scenario of a simple paraelectric-
ferroelectric phase transition (scenario A).

{\it Critical behavior of the order parameter.} There has been
considerable interest in theoretical understanding of the critical
behavior of uniaxial ferroelectrics near the TCLP and many
previous studies analyzed phase transition anomalies of
Sn$_2$P$_2$S$_6$ and derived solid solutions with the hope to
determine the critical exponents relevant to its unconventional
universality class. Since the $T-p$ phase diagram itself is far
from being established, we did not try to enter into the problem
of the critical phenomena. Nevertheless, for the sake of
comparison with other studies, we have analyzed our
order-parameter data with the usually assumed power law dependence
on the reduced temperature $\tau=(T_{\rm C}-T)/T_{\rm C}$.
Interestingly, the critical exponent $\beta$ obtained from the
adjustment of the formula (\ref{intensity}) with $\eta \propto
\tau^\beta$ to the experimental data
 (see Fig.\,\ref{newLabelF3})  was substantially lower than the
mean-field value of 0.5. This result is in contradiction with the
analysis of Refs.\,\onlinecite{Guranich01} and
\onlinecite{Guranich05}. On the other hand, our values of $\beta$
  measured above $p_{\rm TCP}$ = 0.18\,GPa are nicely joining those obtained below
$p_{\rm TCP}$
from the spontaneous polarization (respectively pyroelectric
coefficient) data of
Sn$_2$P$_2$(Se$_{0.04}$S$_{0.96}$)$_6$.\cite{Sli90} Moreover, by
far the best fit [see Fig.\,\ref{newLabelF3}(a)] was obtained for
adjustment to the expected logarithm-corrected power
law\cite{Larkin69,Folk93,Aharony73}
\begin{eqnarray}
\eta = A\tau^\beta \left|  {\rm ln} \left(\tau\right)\right|^x,
\label{order parameter}
\end{eqnarray}
with $\beta=1/4$ and $x=1/10$, that is, values imposed
   according to the predictions for the critical behavior in the vicinity of a TCLP of an uniaxial ferroelectric\cite{Folk93} (see Table II.).


\section{Conclusion}

In conclusion, this paper clarifies the nature of the
ferroelectric phase transition in the Sn$_2$P$_2$S$_6$ single
crystal under a moderate hydrostatic pressure (up to 0.6\,GPa).
The neutron diffraction experiments indicate clearly that the
ferroelectric order parameter vanishes in a fairly continuous
manner at a well-resolved transition temperature. In spite of
considerable efforts, no evidence of an intermediate
incommensurate phase was obtained. While we cannot fully exclude
the existence of an extremely long-wavelength modulation below the
phase transition temperature, the most simple explanation of our
observations is that at least up to 0.6\,GPa, there is only a
simple paraelectric-ferroelectric phase transition in
Sn$_2$P$_2$S$_6$. Still, the temperature dependence of the order
parameter happens to be best reproduced by the logarithm-corrected
power law expected for a uniaxial ferroelectric system in the
vicinity of a TCLP. Thus, it is quite possible that the
thermodynamical properties of the Sn$_2$P$_2$S$_6$ compound and
derived solid solutions may reflect the theoretical closeness to
the ideal TCLP behavior and so it should be taken into account in
analysis of the data.

\section*{Acknowledgments}

The work was partly supported by the Grant Agency of the Czech
Republic (Project No. P204/10/0616 and No. 202/09/H041) and by
project No. SVV-2012-265307. Y.V. is grateful for the support
during the visit to LLB, and all authors thank Fran\c{c}ois
Maignen, St\'{e}phane Pailh\`{e}s and Patrick Baroni for technical
supports at LLB. K.Z.R. gratefully acknowledges past support from
the Alexander von Humboldt Foundation, Germany, and current
support from the Young Investigators Group Programme of the
Helmholtz Association, Germany, contract VH-NG-409.



\begin{thebibliography}{99}


\bibitem{lees12} S. Lee, J. A. Bock, S. Trolier-McKinstry and C. A. Randall,
J. Eur. Ceram. Soc. {\bf 32}, 3971 (2012).

\bibitem{seid10}
S. Y. Yang, J. Seidel, S. J. Byrnes, P. Shafer, C.-H. Yang, M. D.
Rossell, P. Yu, Y.-H. Chu, J. F. Scott, J. W. Ager III, L. W.
Martin and R. Ramesh,
 Nat. Nanotechnol. {\bf 5}, 143 (2010).

\bibitem{seid12} J. Seidel, D. Fu, S.-Y. Yang, E. Alarc\'{o}n-Llad\'{o}, J. Wu, R. Ramesh
and J. W. Ager III,
Phys. Rev. Lett. {\bf 107}, 126805 (2011).


\bibitem{benn12A} J. W. Bennett, K. F. Garrity, K. M. Rabe  and D. Vanderbilt, Phys.
Rev. Lett. {\bf 109}, 167602 (2012).

\bibitem{dand12} D. Daranciang, M. J. Highland, H. Wen, S. M. Young, N. C. Brandt,
H. Y. Hwang, M. Vattilana, M. Nicoul, F. Quirin, J. Goodfellow, T. Qi, I. Grinberg,
D. M. Fritz, M. Cammarata, D. Zhu, H. T. Lemke, D. A. Walko, E. M. Dufresne,
Y. Li, J. Larsson, D. A. Reis, K. Sokolowski-Tinten, K. A. Nelson, A. M. Rappe,
P. H. Fuoss, G. B. Stephenson  and A. M. Lindenberg,
Phys. Rev. Lett. {\bf 108}, 087601 (2012).

\bibitem{imla11} M. Imlau, V. Dieckmann, H. Badorreck  and A. Shumelyuk, Opt. Mater. Express {\bf 1}, 953 (2011).

\bibitem{tiwa09} D. Tiwari and S. Dunn, J. Mater.Sci. {\bf 44}, 5063 (2009).

\bibitem{benn12B} J. W. Bennett and K. M. Rabe, J. Solid State Chem. {\bf 195}, 21 (2012).

\bibitem{book} Yu. M. Vysochanskii, T. Janssen, R. Currat, R. Folk, J. Banys,
J. Grigas  and V. Samulionis, Phase Transitions in Phosphorous
Chalcogenide Ferroelectrics (Vilnius University, Vilnius, 2006).

\bibitem{Levanyuk}
R. Blinc and A. P. Levanyuk, {\em Incommensurate Phases in
Dielectrics, Vol. {\bf 1}} (Amsterdam: North-Holand, 1986).

\bibitem{EijtSnPS}
S. W. H. Eijt, R. Currat, J. E. Lorenzo, P. Saint-Gregoire, B. Hennion  and Yu. M. Vysochanskii,
Eur. Phys. J. B {\bf 5}, 169 (1998).

\bibitem{Diehl}
H. W. Diehl, M. A. Shpot  and R. K. P. Zia,
Phys. Rev. B {\bf 68}, 224415 (2003).


\bibitem{Kosta07}
K. Z. Rushchanskii, Yu. M. Vysochanskii  and D. Strauch,
Phys. Rev. Lett. {\bf 99}, 207601 (2007).


\bibitem{Kohutych}
A. Kohutych, R. Yevych, S. Perechinskii, V. Samulionis, J. Banys  and Yu. M. Vysochanskii,
Phys. Rev. B {\bf 82}, 054101 (2010).


\bibitem{Oleaga} A. Oleaga, A. Salazar, A. A. Kohutych  and Yu. M.
Vysochanskii, J. Phys.: Condens. Matter {\bf 23}, 025902 (2011).

\bibitem{Korda} V. Yu. Korda, S. V. Berezovsky, A. S. Molev, L. P.
Korda  and V. F. Klepikov, Physica B {\bf 407}, 3388 (2012).

\bibitem{moro10} A. N. Morozovska, E. A. Eliseev, J. J. Wang, G. S. Svechnikov, Yu. M.
Vysochanskii, V. Gopalan and L.-Q. Chen, Phys. Rev. B {\bf 81}, 195437 (2010).


\bibitem{AsiGom} A. V. Gomonnai, A. A. Grabar, Y. M. Vysochanskii, A. D. Belyayev,
V. F. Machulin, M. I. Gurzan  and V. Y. Slivka,
Fiz. Tverd. Tela (Leningrad) {\bf 23}, 3602 (1981), in Russian.

\bibitem{Carpentier}
C. D. Carpentier and R. Nitsche,
Mater. Res. Bull. {\bf 9}, 1097 (1974).

\bibitem{Barsamian86}
T. K. Barsamian, S. S. Khasanov, V. Sh. Shekhtman, Yu. M.
Vysochanskii  and V. Yu. Slivka,
Ferroelectrics {\bf 67}, 47 (1986).

\bibitem{Israel}
R. Israel, R. de Gelder, J. M. M. Smits, P. T. Beurskens, S. W. H.
Eijt, Th. Rasing, H. van Kempen, M. M. Maior and S. F. Motrija,
Z. Kristall. {\bf 213}, 34 (1998).


\bibitem{Folk93}
R. Folk and G. Moser,
Phys. Rev. B {\bf 47}, 13992 (1993).

\bibitem{Folk99}
R. Folk, Phase Trans. {\bf 67}, 645 (1999).

\bibitem{VysochJETFbis}
Yu. M. Vysochanskii, V. G. Furtsev, M. M. Khoma, A. A. Grabar, M.
I. Gurzan, M. M. Mayor, S. I. Perechinskii, V. M. Rizak and V. Yu.
Slivka,
Zh. Eksp. Teor. Fiz. {\bf 91}, 1384 (1986) [Sov. Phys. JETP {\bf 64}, 816 (1986)].

\bibitem{TCLP}
Yu. M. Vysochanskii and V. Yu. Slivka,
Sov. Phys. Usp. {\bf 35}, 123 (1992).

\bibitem{Anderson09}
O. Andersson, O. Chobal, I. Rizak  and V. Rizak, Phys. Rev. B {\bf
80}, 174107 (2009).

\bibitem{Sli90}
A. G. Slivka, E. I. Gerzanich, P. P. Guranich  and V. S. Shusta,
Ferroelectrics {\bf 103}, 71 (1990).

\bibitem{Sli99}
A. G. Slivka, E. I. Gerzanich, P. P. Guranich, V. S. Shusta  and V. M. Kedyulich,
Condens. Matter Phys. {\bf 2}, 415 (1999).

\bibitem{Guranich01}
P. P. Guranich, R. V. Kabal, A. G. Slivka  and E. I. Gerzanich,
Ukr. J. Phys. Opt. {\bf 2}, 179 (2001).

\bibitem{Guranich05}
P. P. Guranich, A. G. Slivka, V. S. Shusta, O. I. Gerzanich  and I. Yu. Kuritsa,
Ferroelectrics {\bf 316}, 177 (2005).

\bibitem{Gerzanich08} E. I. Gerzanich,  Ukr. J. Phys. Opt. {\bf 9}, 129 (2008).

\bibitem{look} See Fig.\,1. in Ref.\,\onlinecite{Sli90}.

\bibitem{Zapeka11}
B. Zapeka, O. Mys  and R. Vlokh, Ferroelectrics {\bf 418}, 143 (2011).

\bibitem{Vysoch89}
Yu. M. Vysochanskii, M. M. Mayor, V. M. Rizak, V. Yu. Slivka and
M. M. Khoma, Zh. Eksp. Teor. Fiz. {\bf 95}, 1355 (1989) [Sov.
Phys. JETP {\bf 68}, 782 (1989)].

\bibitem{Vysoch10}
Yu. M. Vysochanskii, A. A. Kohutych, A. V. Kityk, A. V. Zadorozhna, M. M. Khoma and A. A. Grabar,
Ferroelectrics {\bf 399}, 83 (2010).

\bibitem{ferro}
G. Dittmar and H. Schafer,
Z. Naturforsch. B {\bf 29}, 312 (1974).

\bibitem{para}
B. Scott, M. Pressprich, R. D. Willet and D. A. Cleary,
J. Solid State Chem. {\bf 96}, 294 (1992).

\bibitem{EijtSnPSe}
S. W. H. Eijt, R. Currat, J. E. Lorenzo, P. Saint-Gregoire, S.
Katano, T. Janssen, B. Hennion and Yu. M. Vysochanskii,
J. Phys.: Condens. Matter {\bf 10}, 4811 (1998).

\bibitem{Barsamian93}
T. K. Barsamian, S. S. Khasanov and V. Sh. Shekhtman,
Ferroelectrics {\bf 138}, 63 (1993).

\bibitem{JANA}
V. Petricek, M. Dusek and L. Palatinus, Jana2006, the
crystallographic computing system. Institute of Physics AS CR,
Prague, Czech Republic (2006).

\bibitem{Hlinka05}
J. Hlinka, R. Currat, M. de Boissieu, F. Livet and Yu. M. Vysochanskii,
Phys. Rev. B {\bf 71}, 052102 (2005).

\bibitem{Slivka05}  A. G. Slivka, V. M. Kedyulich and E. I. Gerzanich, Ferroelectrics {\bf 317}, 89
(2005).


\bibitem{Tyagur}
Yu. Tyagur, E. I. Gerzanich and A. G. Slivka,
Sov. Phys. J. {\bf 28}, 739 (1985).

\bibitem{Anderson11}
O. Andersson, O. Chobal, I. Rizak, V. Rizak and V. Sabadosh,
Phys. Rev. B {\bf 83}, 134121 (2011).

\bibitem{Say}
A. Say, O. Mys, D. Adamenko, A. Grabar, Y. Vysochanskii, A. Kityk and R. Vlokh,
Phase Trans. {\bf 83}, 123 (2010).

\bibitem{UFZ1} Yu. M. Vysochanskii, S. F. Motrja, S. I. Perechinskii, V. M. Rizak, I. M. Rizak and V. Yu. Slivka, Ukr. fiz. Zh. {\bf 36}, 728 (1991).

\bibitem{UFZ2} See Ref.\,\onlinecite{book}, Fig. 3.25, page 167.

\bibitem{UFZ3} I. M. Rizak, V. M. Rizak, S. I. Perechinskii, Yu. M. Vysochanskii and V. Yu.
Slivka, Ferroelectrics {\bf 143}, 67 (1993).

\bibitem{HlinkaBCCD}
J. Hlinka, M. Iwata and Y. Ishibashi,
J. Phys. Soc. Jpn. {\bf 68}, 126 (1999).

\bibitem{VysochJETF}
Yu. M. Vysochanskii, V. G. Furtsev, M. M. Khoma, M. I. Gurzan and V. Yu. Slivka,
Zh. Eksp. Teor. Fiz. {\bf 89}, 939 (1985) [Sov. Phys. JETP {\bf 62}, 540 (1985)].

\bibitem{Larkin69}
A. I. Larkin and D. E. Khmel'nitskii,
Zh. Eksp. Teor. Fiz. {\bf 56}, 2087 (1969) [
Sov. Phys. JETP {\bf 29}, 1123 (1969)].

\bibitem{Aharony73}
A. Aharony,
Phys. Rev. B {\bf 8}, 3363 (1973).

\bibitem{thiourea}
J. Hlinka, J. Petzelt,  B. Brezina and R. Currat, Phys. Rev. B {\bf 66}, 132302 (2002).

\bibitem{Currat}
R. Currat and T. Janssen, Solid State Phys. {\bf 41}, 201 (1988).

\bibitem{Dvorak} J. Hlinka, T. Janssen and V. Dvorak, J. Phys.: Condens. Matter {\bf 11}, 3209 (1999).

\bibitem{Padlewski} S. Padlewski, A. K. Evans, C. Ayling and V. Heine, J. Phys.: Condens. Matter {\bf 4}, 4895 (1992).

\bibitem{Volkov} A. A. Volkov, G. V. Kozlov, N. I. Afanas'eva, Yu. M. Vysochanskii,
A. A. Grabar and V. Yu. Slivka, Sov. Phys. Solid State {\bf 25}, 1482 (1983).


\end{thebibliography}
\end{document}